\documentclass[twocolumn]{aastex631}

\newcommand{\obj}{J0725$+$5835}

\received{XXX}
\revised{XXX}
\accepted{XXX}

\submitjournal{ApJ}

\shorttitle{XRG J0725$+$5835}
\shortauthors{Yang et al.}
\begin{document}

%\title{Template \aastex Article with Examples: 
%v6.31\footnote{Released on March, 1st, 2021}}

%\title{Impulsive gravitational interaction as the origin of X-shaped jet in J0725$+$5835?}

\title{The X-shaped radio galaxy J0725$+$5835 is associated with an AGN pair}

\correspondingauthor{Xiaolong Yang}
\email{yangxl@shao.ac.cn}

\author[0000-0002-4439-5580]{Xiaolong Yang}
\affiliation{Shanghai Astronomical Observatory, Key Laboratory of Radio Astronomy, Chinese Academy of Sciences, Shanghai 200030, China}
\affiliation{Shanghai Key Laboratory of Space Navigation and Positioning Techniques, Shanghai Astronomical Observatory, Chinese Academy of Sciences, Shanghai 200030, China}
\affiliation{Kavli Institute for Astronomy and Astrophysics, Peking University, Beijing 100871, China}

\author[0000-0003-2354-1574]{Jialu Ji}
\affiliation{Glasgow of Strathclyde, 16 Richmond Street, Glasgow G11XQ, UK}
\affiliation{Shanghai Normal University, No.100 Guilin Rd. Shanghai 200234, China}

\author[0000-0002-5535-4186]{Ravi Joshi}
\affiliation{Indian Institute of Astrophysics, Bengaluru, India}
\affiliation{Kavli Institute for Astronomy and Astrophysics, Peking University, Beijing 100871, China}

\author[0000-0002-2322-5232]{Jun Yang}
\affiliation{Department of Space, Earth and Environment, Chalmers University of Technology, Onsala Space Observatory, SE-439\,92 Onsala, Sweden}
\affiliation{Shanghai Astronomical Observatory, Key Laboratory of Radio Astronomy, Chinese Academy of Sciences, Shanghai 200030, China}

\author[0000-0003-4341-0029]{Tao An}
\affiliation{Shanghai Astronomical Observatory, Key Laboratory of Radio Astronomy, Chinese Academy of Sciences, Shanghai 200030, China}

\author[0000-0003-4956-5742]{Ran Wang}
\affiliation{Kavli Institute for Astronomy and Astrophysics, Peking University, Beijing 100871, China}
\affiliation{Department of Astronomy, School of Physics, Peking University, Beijing 100871, China}

\author[0000-0001-6947-5846]{Luis C. Ho}
\affiliation{Kavli Institute for Astronomy and Astrophysics, Peking University, Beijing 100871, China}
\affiliation{Department of Astronomy, School of Physics, Peking University, Beijing 100871, China}

\author{David H. Roberts}
\affiliation{Department of Physics MS-057, Brandeis University, Waltham, MA 02453-0911, USA}

\author{Lakshmi Saripalli}
\affiliation{Raman Research Institute, C. V. Raman Avenue, Sadashivanagar, Bangalore 560080, India}



\begin{abstract}
X-shaped radio galaxies (XRGs) are those that exhibit two pairs of unaligned radio lobes (main radio lobes and "wings"), one of the promising models for the peculiar morphology is jet re-orientation. To clarify it, we conducted the European VLBI Network (EVN) 5\,GHz observation of an XRG \obj, which resembles the archetypal binary AGNs 0402+379 in radio morphology but it is larger in angular size. In our observation, two milliarcsec (mas) scale radio components with non-thermal radio emission are detected, each of them coincides with an optical counterpart with similar photometric redshift and (optical and infrared) magnitude, corresponding to dual active nuclei. Furthermore, with the improved VLA images, we find a bridge between the two radio cores and a jet bending in the region surrounding the companion galaxy, which further supports the interplay between the main and companion galaxies. In addition, we also report the discovery of an arcsec-scale jet in the companion. Given the projected separation of $\sim100$\,kpc between the main and companion galaxies, XRG \obj\ is likely associated with a dual jetted-AGN system. In both EVN and VLA observations, we find signatures that the jet is changing its direction, which is likely responsible for the X-shaped morphology. On the origin of jet re-orientation, several scenarios are discussed.
\end{abstract}



\section{Introduction\label{sec:intro}}

Astronomical observations show that supermassive black holes (SMBHs) are located at the center of most massive galaxies \citep[e.g.,][]{2013ARA&A..51..511K}. According to the hierarchical model of structure formation, galaxy mergers occur frequently in the early Universe \citep{2003ApJ...582..559V}, which naturally results in two SMBHs that sink to the center of a newly formed galaxy due to dynamical friction and form a binary system. Moreover, galaxy mergers can efficiently drive gas inflow to fuel the SMBHs, thus stimulating the active galactic nuclei (AGNs) and nuclear starburst \citep{2006ApJS..163....1H, 2007A&A...468...61D, 2010A&A...518A..56M}. To establish a complete view of galaxy merger evolution, one needs to further explore, e.g. how the strength of star formation, AGN accretion, and feedback evolve with the merger sequence. Even though dual and binary AGNs candidates with separations ranging from $100$\,kpc down to few parsec scales have been detected \citep{2003ApJ...582L..15K, 2008MNRAS.386..105B, 2011ApJ...740L..44F, 2011ApJ...735L..42K, 2011ApJ...737L..19C, 2014MNRAS.437...32W, 2014Natur.511...57D, 2015ApJ...799...72F, 2018A&A...610L...7H, 2017MNRAS.464L..70Y, 2018ApJ...862...29L, 2019ApJ...879L..21G, 2020ApJ...899..154S}, the number of confirmed dual and binary AGNs is still smaller than predicted \citep{2018RaSc...53.1211A}, and one reason for this deficiency is the lack of effective candidates. Here we define dual AGNs as gravitational interacting systems with a separation smaller than $\approx 100$\,kpc.

Under suitable conditions, an AGN can launch a pair of relativistic jets of non-thermal radio emission with dimensions that can evolve up to megaparsecs \citep[e.g.,][]{2014ARA&A..52..589H, 2017MNRAS.469.2886D}. \citet{1978Natur.276..588E} noted that a symmetric jet distortion in galaxy NGC~326 can be caused by jet precession, while further studies indicate that a more rapid change of jet direction is required to form a nearly orthogonal distortion in NGC~326 \citep[e.g.][]{2001A&A...380..102M} and other morphologically similar sources \citep[i.e. X-shaped radio galaxies, XRGs, ][]{2002MNRAS.330..609D}. There are two scenarios concerning the jet re-orientation: jet precession due to binary black hole evolution \citep{2007MNRAS.377.1215Z, 2007ApJ...661L.147B, 2019MNRAS.482..240K, 2020MNRAS.499.5765H} and the sudden spin-flip after binary black hole coalescence \citep{2002Sci...297.1310M}. On the other hand, backflowing plasma from the jet may also be warped as the "wing-shape" in XRGs \citep{2017A&A...606A..57R, 2020MNRAS.495.1271C}.

This work focuses on the XRGs and their origin is still debated \citep[see][and references therein]{2019ApJS..245...17Y}. The jet precession model requires that binary black holes are located in parsec-scale \citep{2019MNRAS.482..240K}, while it lacks observational support. Interestingly, we find observational evidence that several XRGs are in kilo-parsec-scale galaxy pairs \citep[e.g.][]{1980A&A....85..101B, 1984MNRAS.210..929L, 2001A&A...380..102M, 2006ApJS..164..307M}. A study of $\sim$100 radio sources associated with dumbbell galaxies, i.e., two nearly equally bright elliptical galaxies within a common envelope, has revealed remarkably distorted radio structures in about a dozen of them \citep{1982AJ.....87..602W}. Currently, there is enough observational evidence to show that galaxy pairs are associated with jet distortion \citep[e.g.][]{1991AJ....102.1960P, 1993ApJ...416..157B, 2019MNRAS.488.3416H}, therefore, it occasionally forms X-shaped structures.

\begin{figure}
\centering \includegraphics[scale=0.36]{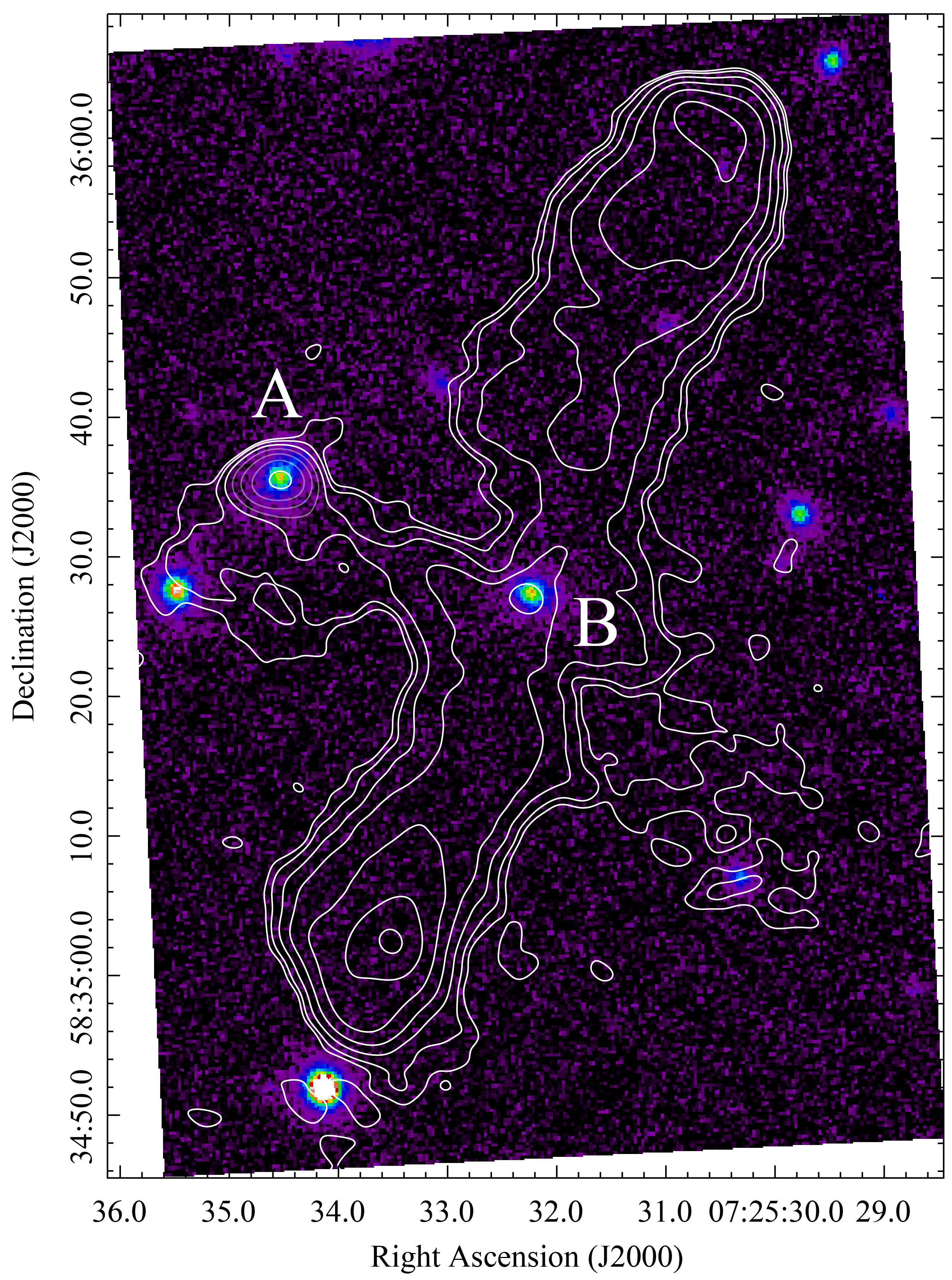}
\caption{Radio contours (white lines) of \obj\ overlaid on the Pan-STARRS1 $i$-band image (pseudo-color image). The radio contours are from the naturally-weighted VLA S-band (3\,GHz) image that was obtained on 2016-May-28 with the B-array. The contours increase in steps of $(1.25, 2, 4, 8, ...)\times0.04\,\mathrm{mJy\,beam^{-1}}$. \label{fig:opt}}
\end{figure}

\obj\ was selected as an X-shaped radio galaxy \citep{2007AJ....133.2097C} because the directions of its primary and secondary jets are almost orthogonal. What caught our attention is its similar morphology with the parsec-scale binary AGN 0402+379 \citep{2006ApJ...646...49R}. There is a radio core at the symmetric center of \obj\ driving the large-scale radio jet, while a compact (at arcsec-scales) radio component to its north-east direction is likely a massive neighboring galaxy that could affect the jet shape of the main radio galaxy. In the following, we refer to the radio core of the primary jet as core B and the compact radio component in the north-east direction as core A (see Figure \ref{fig:opt}); both core A and core B have Panoramic Survey Telescope and Rapid Response System (Pan-STARRS) counterparts (see Pan-STARRS1 $i$-band image in Figure \ref{fig:opt}) and show similar photometric redshift $z\sim0.42$ (see Section \ref{sec:dual-agn}). In this paper, we present the results of observations of the XRG \obj\ by the European Very Long Baseline Interferometry (VLBI) Network (EVN) and the Jansky Very Large Array (VLA). 

This paper is organized as follows: we describe the EVN and VLA observations and the data reduction in Section \ref{sec:obs} and present the imaging results of \obj\ in Section \ref{sec:result}. We discuss the interplay between cores A and B, jet re-orientation, and simultaneous scenarios in Section \ref{sec:dis}. Throughout the paper, a standard $\Lambda$CDM cosmology is used with $H_0=\mathrm{71\,km\,s^{-1}\,Mpc^{-1}}$, $\Omega_m=0.27$, and $\Omega_\Lambda=0.73$.

\begin{figure*}
\centering \includegraphics[scale=0.3]{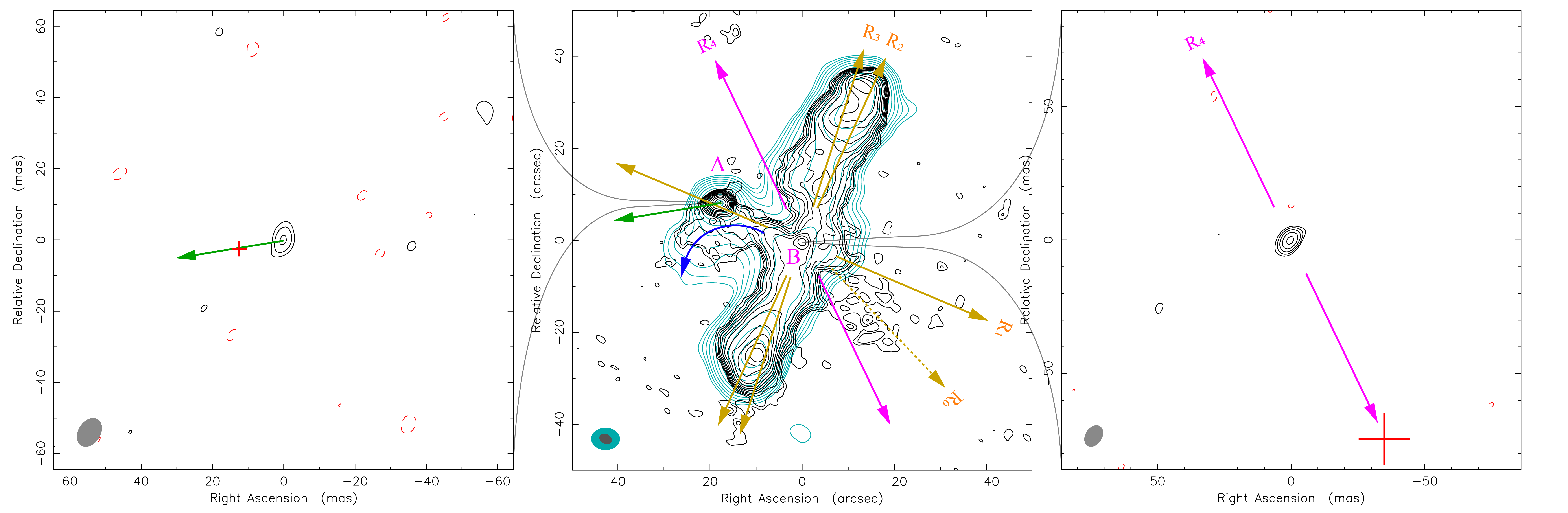}
\caption{Left and right panel: Naturally-weighted EVN 5\,GHz image of cores A and B, respectively; Middle panel: overlapped contours between VLA B-array L- (cyan) and S-band (black). The corresponding beam is shown in the lower-left corner of each panel. The red crosses in the left and right panels show the astrometric position obtained from the VLA B-array S-band data (see Table \ref{tab:evn+vla}). Where the arrows are defined as $R_0$: the direction of fossil radio emission, $R_1$: the wings, $R_2$: the direction of lobes, $R_3$: the direction from core to the tip of hotspots, and the green arrow and $R_4$: the arrows are obtained from left and right panel, respectively, indicate the direction from VLBI to VLA position. \label{fig:model}}
\end{figure*}

\begin{deluxetable*}{cccccccccc}
\tablenum{1}
\tablecaption{Summary of the VLA and EVN imaging and circular Gaussian fitting results of \obj.\label{tab:evn+vla}}
\tablewidth{0pt}
\tablehead{
\colhead{Facility} &\colhead{$\nu$} & \colhead{$\alpha$}  & \colhead{$\sigma_\alpha$} & \colhead{$\delta$}   & \colhead{$\sigma_\delta$}    & \colhead{$S_i$}       & \colhead{$\theta$} & \colhead{$logT_b$} & \colhead{$S_p$} \\
\colhead{}         &\colhead{(GHz)} & \colhead{(J2000)}   & \colhead{(mas)}           & \colhead{(J2000)}    & \colhead{(mas)}              & \colhead{(mJy)}   & \colhead{(mas)}     & \colhead{(K)}      & \colhead{(mJy\,beam$^{-1}$)}
}
\decimalcolnumbers
\startdata
Core A \\
\hline
VLA A-array & 1.5 &   $\mathrm{07^h25^m34^s.5090}$     & 7.27 & $\mathrm{+58^\circ35^{\prime}35^{\prime\prime}.699}$     & 7.27   &$3.865\pm0.178$ &        &              &  $3.590\pm0.198$   \\
VLA B-array & 3   &   $\mathrm{07^h25^m34^s.5220}$     & 1.88 & $\mathrm{+58^\circ35^{\prime}35^{\prime\prime}.499}$     & 1.88   &$2.822\pm0.036$  &        &              &  $3.050\pm0.027$     \\
EVN          & 5   &   $\mathrm{07^h25^m34^s.520369}$   & 0.34 & $\mathrm{+58^\circ35^{\prime}35^{\prime\prime}.50125}$   & 0.36   &$0.076\pm0.021$ &$\lesssim1.84$ &$\gtrsim6.37$  &$0.084\pm0.024$     \\
\hline
Core B \\
\hline
VLA A-array & 1.5 &   $\mathrm{07^h25^m32^s.2700}$     & 62   & $\mathrm{+58^\circ35^{\prime}27^{\prime\prime}.100}$     & 62     &$0.973\pm0.174$  &        &              & $0.690\pm0.198$    \\
VLA B-array & 3   &   $\mathrm{07^h25^m32^s.2571}$     & 9.28 & $\mathrm{+58^\circ35^{\prime}27^{\prime\prime}.100}$     & 9.28   &$1.196\pm0.026$  &        &              & $0.913\pm0.027$    \\
EVN          & 5   &   $\mathrm{07^h25^m32^s.261525}$   & 0.19 & $\mathrm{+58^\circ35^{\prime}27^{\prime\prime}.17425}$   & 0.15   &$0.227\pm0.033$   &$\lesssim1.23$ &$\gtrsim7.28$  &$0.280\pm0.038$  \\
\enddata

\tablecomments{Columns give (1) source name/facilities, (2) frequency, 
(3–6) J2000 positions and errors, (7) total flux density, (8) source angular size, (9) brightness temperature, and (10) peak brightness. }

\end{deluxetable*}

\section{Observations and data reduction} \label{sec:obs}

\subsection{VLBI observation and data reduction}
We observed \obj\ on 2017 November 14 with ten antennas of the EVN at C-band ($5$\,GHz, the project code RSY06, PI: Xiaolong Yang). The total observation time is 2 h with a data recording rate of 2\,Gbps. The observation was performed in phase-referencing mode, using 0724$+$571 (R.A. $=\mathrm{07^{h}28^{m}49^{s}.6317}$, DEC. $=57^{\circ}01^{\prime}24^{\prime\prime}.374$) as the phase reference calibrator. 

We calibrated the EVN data in the Astronomical Image Processing System (AIPS), a software package developed by the National Radio Astronomy Observatory (NRAO) of U.S. \citep{2003ASSL..285..109G}, following the standard procedure. Apriori amplitude calibration was performed using the system temperatures and antenna gain curves provided by each station. The Earth orientation parameters were obtained and calibrated using the measurements from the U.S. Naval Observatory database, and the ionospheric dispersive delays were corrected from a map of the total electron content provided by the Crustal Dynamics Data Information System (CDDIS) of NASA \footnote{\url{https://cddis.nasa.gov}}. The opacity and parallactic angles were also corrected using the auxiliary files attached to the data. The delay in the visibility phase was solved using the phase reference calibrator 0724$+$571. A global fringe-fitting on the phase-referencing calibrator 0724$+$571 was performed by taking the calibrator's model to solve miscellaneous phase delays of the target. The target source's data were exported to DIFMAP \citep{1997ASPC..125...77S} for self-calibration and model fitting. Especially, in our VLBI data reduction, we take the radio coordinate from the VLA B-array S-band (see Section \ref{subsec:vla}) as a reference for core A, which proved to be sufficiently accurate (see Table \ref{tab:evn+vla}). No self-calibration was applied to the target source since it is too weak ($\sim7\sigma$ and $13\sigma$ for cores A and B, respectively). The final image was created using natural weighting; see Figure \ref{fig:model}. 

\subsection{VLA observation and data reduction} \label{subsec:vla}
This source is a part of the VLA L- (1.4\,GHz) and S-band (3\,GHz) surveys of 89 XRGs \citep{2018ApJ...852...47R}. In this work, we have retrieved two VLA B-array datasets at L- and S-band (project code 16A-220, PI: Lakshmi Saripalli) and one VLA A-array dataset at L-band (project code 16B-023, PI: Lakshmi Saripalli) from NRAO Science Data Archive \footnote{\url{https://archive.nrao.edu/archive/advquery.jsp}}. The VLA B-array L-band observation was performed on 2016 May 29 with a bandwidth of 1~GHz and a total on-source time of 4.82\,min, while the S-band observation was made on 2016 May 28 with a bandwidth of 2 GHz and a total on-source time of 4.74\,min. The VLA A-array L-band observation was performed on 2016 December 27 with a bandwidth of 1 GHz and a total on-source time of 3.57\,min. All observations 
were scheduled in a full polarization mode, and used J0713+4349 (R.A. $=\mathrm{07^{h}13^{m}38^{s}.164}$, DEC. $=+43^{\circ}49^{\prime}17^{\prime\prime}.21$) as the secondary flux density calibrator. Although the data have been published by \citet{2018ApJ...852...47R} and \citet{2018ApJ...852...48S}, in order to ensure uniformity in the analysis across all the parameters we have re-reduced the data using the Common Astronomy Software Application \citep[CASA v5.1.1,][]{2007ASPC..376..127M}. 

Our data analysis followed the standard routines described in the CASA Cookbook \footnote{\url{https://casa.nrao.edu/casadocs}}. We used 3C\,138 and 3C\,147 as the primary flux calibrators for VLA B-array and A-array datasets, respectively, and adopted the flux density standards `Perley-Butler 2017' \citep{2017ApJS..230....7P} to set the overall flux density scale; after that, we bootstrapped to the secondary flux density calibrator (J0713+4349) and the target. Antenna delay and bandpass corrections were also determined by fringe-fitting the visibilities. Furthermore, J0713+4349 was also used as a phase calibrator in the observations. We determined the complex gains from the phase calibrator and applied it to the target. We also performed an ionospheric correction using the data obtained from the CDDIS archive. Deconvolution, self-calibration, and model fitting were performed in the DIFMAP software package. For the good uv-coverage of the VLA observations and the high signal-to-noise ratio (SNR$>$9), the VLA data allowed self-calibration. This was initially performed only on phase, and subsequently on both phase and amplitude when we achieved a good model. Natural weighting was used to create the final image.

\begin{figure*}
\centering \includegraphics[scale=0.34]{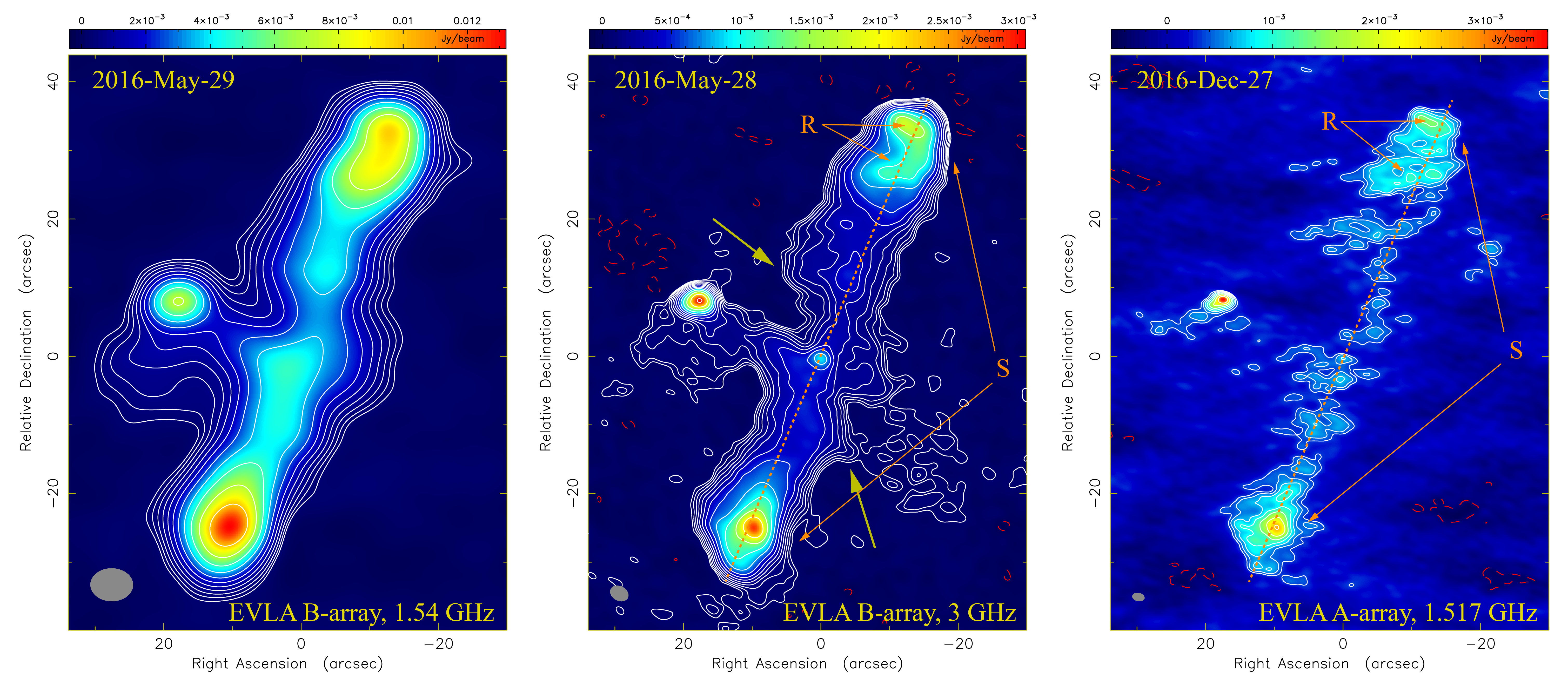}
\caption{Naturally-weighted VLA images of \obj. Left panel: L-band (1.54\,GHz) image obtained on 2016-May-29 with the B-array; Middle panel: S-band (3\,GHz) image obtained on 2016-May-28 with the B-array; Right panel: L-band (1.51\,GHz) image obtained on 2016-Dec-27 with the A-array. The contours increase in steps of $(-1, 1, 1.41, 2, 2.83, 4, ...)\times3\sigma$, where the $1\sigma$ root-mean-square noise levels are $0.103$, $0.015$, and $0.110\,\mathrm{mJy\,beam^{-1}}$ for the left, middle, and right panels, respectively. In the lower-left corner of each panel, the gray ellipse illustrates the synthesized beam for each band. The corresponding radio flux density scale is linked with a color map in each color bar. R: Wide terminal hotspots and ring-like structure on the northern side, and the southern hotspot is not at the edge of its lobe. S: Northern hotspot on the opposite of lobe as the southern hotspot. The yellow arrows show the bumps. \label{fig:vla}}
\end{figure*}

\section{Results} \label{sec:result}

In Figure \ref{fig:vla}, we show the VLA B-array images obtained at L- and S-band and the VLA A-array image at L-band. In the B-array L-band observation, we obtained a synthesized beam size of $6.24\arcsec\times4.83\arcsec$ at a position angle of $89.6^\circ$, while in the A-array L-band observation, we obtained a synthesized beam size of $1.83\arcsec\times1.21\arcsec$ at a position angle of $78.1^\circ$. In the VLA B-array S-band observation, we obtained a synthesized beam size of $2.83\arcsec\times2.13\arcsec$ at a position angle of $61.5^\circ$. We perform a two-dimensional Gaussian model-fitting for the VLA A-array L-band and B-array S-band data, and the results are listed in Table \ref{tab:evn+vla}, where the uncertainties of the integrated flux density and peak flux density are estimated using the method described by \citet{2020ApJ...904..200Y}. In our EVN observation, both core A and core B were detected at 4.92\,GHz (see Figure \ref{fig:model}). The peak flux densities of A and B are $0.084\pm0.024$ and $0.280\pm0.038\,\mathrm{mJy\,beam^{-1}}$, yielding an SNR of $6.8$ and $13.3$, respectively. For the EVN data, two-dimensional Gaussian models were used to fit the visibility data for each target to obtain the integrated flux density and size of the Gaussian components (full width at half maximum, FWHM). The model-fitting results are listed in Table \ref{tab:evn+vla}. For the VLA data, we fit two-dimensional Gaussian models to cores A and B only to obtain their integrated flux densities of $3.865\pm0.178$ and $0.973\pm0.174$\,mJy at 1.54\,GHz (VLA A-array), and $2.822\pm0.036$ and $1.196\pm0.026$\,mJy at 3\,GHz (VLA B-array), respectively. By combining the radio flux densities at different frequencies, we can obtain the spectral indices ($S\propto\nu^{+\alpha}$) of core A and core B as $-0.47\pm0.07$ and $+0.30\pm0.27$, respectively.

By fitting two-dimensional Gaussian models to the EVN data, we can estimate the brightness temperatures using the formula \citep[e.g.][]{2005ApJ...621..123U} 
\begin{equation}\label{eq:bt}
T_\mathrm{B}=1.8\times10^9(1+z)\frac{S_i}{\nu^2\theta^2}~\mathrm{(K)},
\end{equation}
where \(S_i\) is the integrated flux density of each Gaussian model component in mJy (column 6 of Table \ref{tab:evn+vla}), $\theta$ is $\mathrm{FWHM}$ of the Gaussian model in mas (column 7 of Table \ref{tab:evn+vla}), \(\nu\) is the observing frequency in GHz, and \(z\) is the redshift; here we used an averaged photometric redshift of $z=0.42$ (see Section \ref{sec:dual-agn}) of core A and core B. The estimated 5\,GHz brightness temperatures are listed in column 8 of Table \ref{tab:evn+vla}. Because the measured component sizes are only upper limits, the radio brightness temperatures should be considered as lower limits. The brightness temperatures for cores A and B are $10^{6.37}$ and $10^{7.28}$\,K, respectively, consistent with a non-thermal origin for the mas-scale radio emission.

In the phase-referenced EVN observation, the astrometric accuracy can be measured by considering the positional uncertainty of phase calibrator 0724$+$571 ($\sigma_p=0.14$\,mas), the errors in phase-transferring from calibrator to target ($\sigma_{pr,\alpha}=0.035$\,mas and $\sigma_{pr,\delta}=0.117$\,mas) \citep{2006A&A...452.1099P} and the rms error of target. The rms error of the target can be estimated as $\theta_\mathrm{FWHM}/2\mathrm{SNR}$, where $\theta_\mathrm{FWHM}$ is the FWHM of the beam and $\mathrm{SNR}$ is the signal-to-noise ratio. While for VLA observations, only the rms astrometric error is considered.

\begin{figure*}
\centering \includegraphics[scale=0.4]{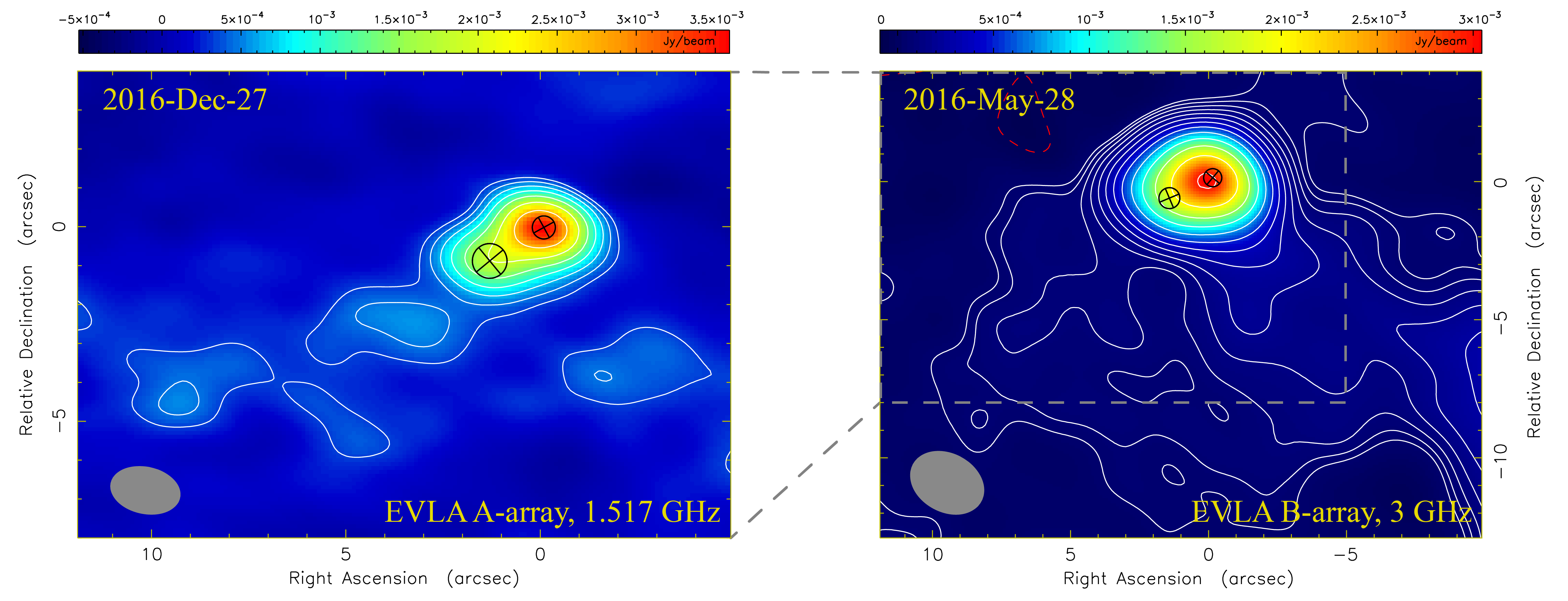}
\caption{VLA images of core A. The left panel shows L-band (1.51\,GHz) image obtained on 2016-Dec-27 with A-array, while the right panel shows S-band (3\,GHz) image obtained on 2016-May-28 with B-array. The contours and root-mean-square noise are the same as the corresponding image in Figure \ref{fig:vla}. In both panels, the black circles represent the Gaussian components. In the lower-left corner of each panel, the grey ellipse illustrates the synthesized beam for each band. The corresponding radio flux density scale is linked with a color map in each color bar. \label{fig:coreA}}
\end{figure*}

\section{Discussion} \label{sec:dis}

\subsection{Interplay between A and B} \label{sec:dual-agn}

\begin{figure}
\centering \includegraphics[scale=0.48]{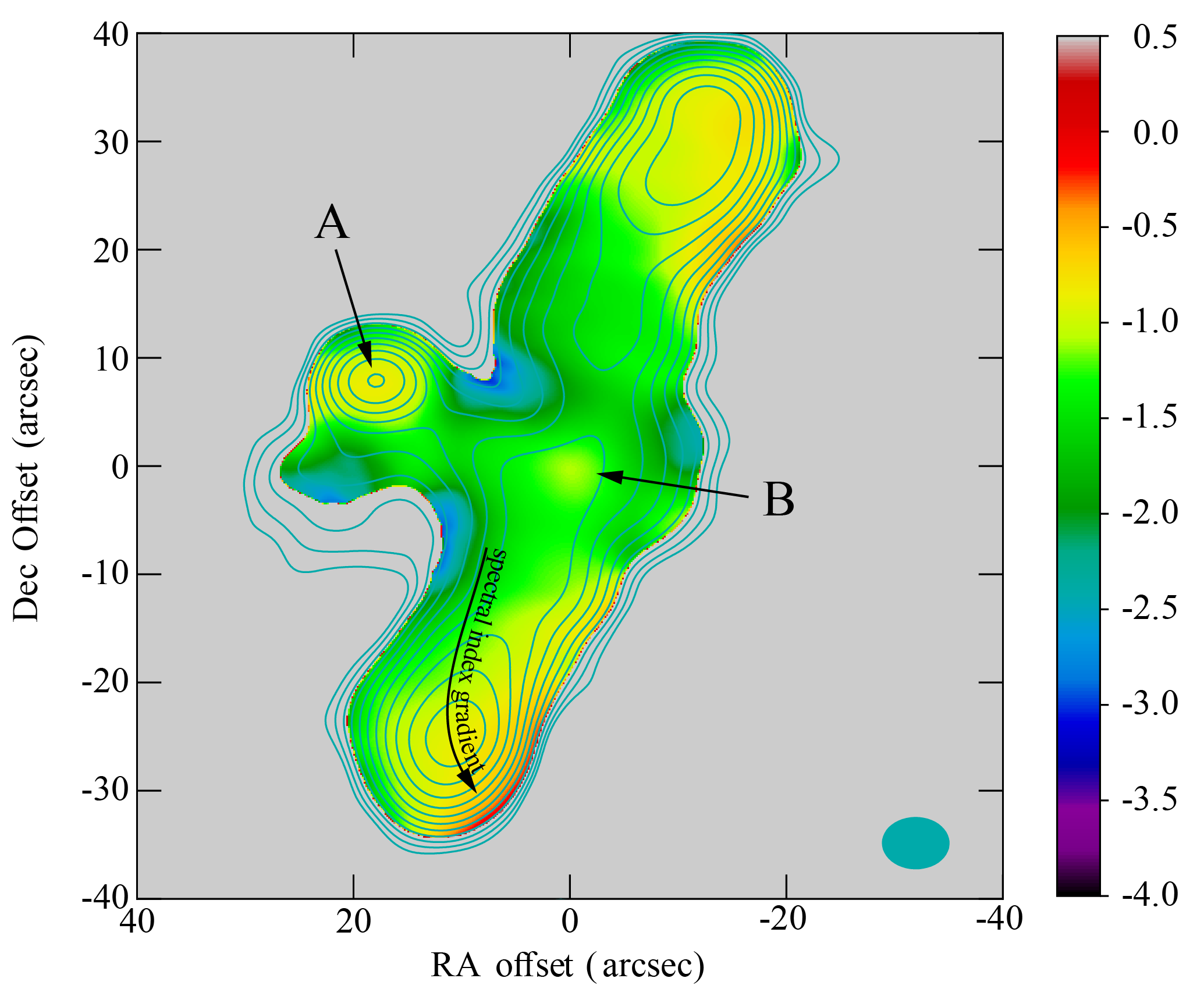}
\caption{Radio spectral index map between 1.5 and 3\,GHz. The spectral index map was created from the VLA B-array S-band and B-array L-band data. The corresponding spectral index scale is linked with a color map in the color bar, where the yellow regions are with the flat spectrum, while the green regions are with the steep spectrum. Here only the regions with radio flux density above $7\sigma$ in both frequencies are presented. The cyan contours are from VLA B-array L-band data and same with the left panel of Figure \ref{fig:vla}, and its beam shows at the lower-right corner with a cyan ellipse. \label{fig:spmap}}
\end{figure}

The EVN detection of compact radio cores with high brightness temperatures ($>10^6\,\mathrm{K}$) and flat radio spectra ($\alpha>-0.5$, spectral index of core A marginally satisfies the criteria, but see Figure \ref{fig:spmap}) supports radio-frequency active AGNs located at both galactic nuclei in A and B. The Dark Energy Camera Legacy Survey \citep[DECaLS, ][]{2019AJ....157..168D} measured a photometric redshift of $0.399\pm0.057$ and $0.442\pm0.028$ for cores A and B, respectively. The similar redshifts between A and B support the possibility that they are in a close pair system.

Since we have no spectroscopic redshift measurements on this source, we checked radio properties to further explore the features supporting an interplay between A and B. Clearly, the radio bridge between cores A and B, as well as the deflection of the eastern radio outflow (jet relics) in the core A region (see the blue arrow in the middle panel of Figure \ref{fig:model}), both hint the interplay between cores A and B. Again, the radio structure around core A is unlikely from itself alone, because an arcsec-scale jet is found in both VLA A-array L-band and B-array S-band images (see Figure \ref{fig:coreA}), and shows a different extension. Additional evidence of the radio bridge between A and B is from the radio spectral index map (Figure \ref{fig:spmap}), which shows the continuous distribution between A and B. Furthermore, the interaction between the radio outflow and core A may also be inferred from the enhanced polarization in the circum-A region \citep[supplementary data of][]{2018ApJ...852...47R}. In the case here, the explanation for the deflection of the radio outflow is straightforward, which could be a combined effect of galactic ram pressure \citep{1984ApJ...281..554F}, density gradients \citep{1984MNRAS.211..767S} or oblique magnetic field \citep{1997ApJ...478...66K, 2021Natur.593...47C} from core A. Since core A is radio-frequency active, it is reasonable to suppose that multiple mechanisms are at play.

According to our EVN 5\,GHz observation, the angular distance between cores A and B is $19.5$\,arcsec. If we take the averaged photometric redshift between cores A and B ($z=0.42$, the scale is $5.504\,\mathrm{kpc\,arcsec^{-1}}$) as a reference, the projected distance between the two AGNs is about 107\,kpc, a marginally dual AGN system \citep{2011ApJ...737..101L}. This also results in a projected distance of $\sim450$\,kpc between the northern and southern hotspots. Cores A and B have nearly equal apparent $r$-band magnitudes of $20.54$ and $20.70$ in the Pan-STARRS sky survey \citep{2016arXiv161205560C}, and there appears to be a tidal-tail structure in the image of galaxy A (see Figure \ref{fig:opt}), which further hints the interplay between two galaxies. The association of an X-shaped jet morphology in interacting galaxies is quite common \citep[e.g.][]{1982AJ.....87..602W, 2001A&A...380..102M, 2007ApJ...664..804M}.

\subsection{Jet is changing its direction?} \label{subsec:reor}
With the EVN and VLA observations, we can measure the coordinates of cores A and B (see Table \ref{tab:evn+vla}). The coordinates measured by the VLA are consistent between A-array L-band and B-array S-band, while the B-array S-band has higher accuracy, so we will use it in the discussion below. At the position of core B, our EVN 5\,GHz measurement has a deviation from the VLA. Interestingly, this deviation is not caused by an error but is intrinsic, i.e. if we take the EVN 5\,GHz location of core B as the reference, then the location of core B measured by the VLA B-array at S-band is at R.A. offset$=-67\pm9$\,mas and DEC offset$=-74\pm9$\,mas (red cross in the right panel of Figure \ref{fig:model}), beyond the $7\sigma$ astrometric error of VLA. Furthermore, it's unlikely to be a systematic error, compared to the VLA coordinates of core A which are close to the EVN measured position (see left panel of Figure \ref{fig:model}).

A reasonable explanation for the position offset is that there is an inner $<10$\,kpc-scale jet in the direction $-124^\circ\pm26^\circ$. Since the inner kpc-scale jet is not resolved by VLA observations, the VLA position of core A is likely an average between the radio core (i.e., the EVN position) and the jet. Such a structure-based peak-shift along the jet is common in AGNs \citep[e.g.][]{2017A&ARv..25....4B, 2021ApJ...911L..11E}. Furthermore, a frequency-based core shift may contribute to a few mas-scale positional shifts \citep{2012ApJ...756..161S}. The frequency-based core shift is along with the jet, so it will not affect the hypothesis of an inner jet. Though other scenarios can not be ruled out, e.g. an off-nuclear AGN or a radio bright star-forming region in the circumnuclear region of galaxy B, it seems like the jet re-orientation is self-consistent with other signatures below. We marked the direction from the EVN to VLA positions of core B as $R_4$ in the middle and left panels of Figure \ref{fig:model}. It seems the bumps near $R_4$ in the VLA S-band image (see arrows in the middle panel of Figure \ref{fig:vla}) give some hints about the inner jet direction. Indeed, the arcsecond-scale jet direction of core A (Figure \ref{fig:coreA}) is precisely consistent with the direction from the EVN to VLA coordinates (red cross in the left panel of Figure \ref{fig:model}). The inner jet direction of core B is clearly different from the major jet axis direction on arcsec scales (see Figure \ref{fig:model}), thus hinting at an ongoing jet re-orientation at the nuclei.

\begin{figure}
\centering \includegraphics[scale=0.22]{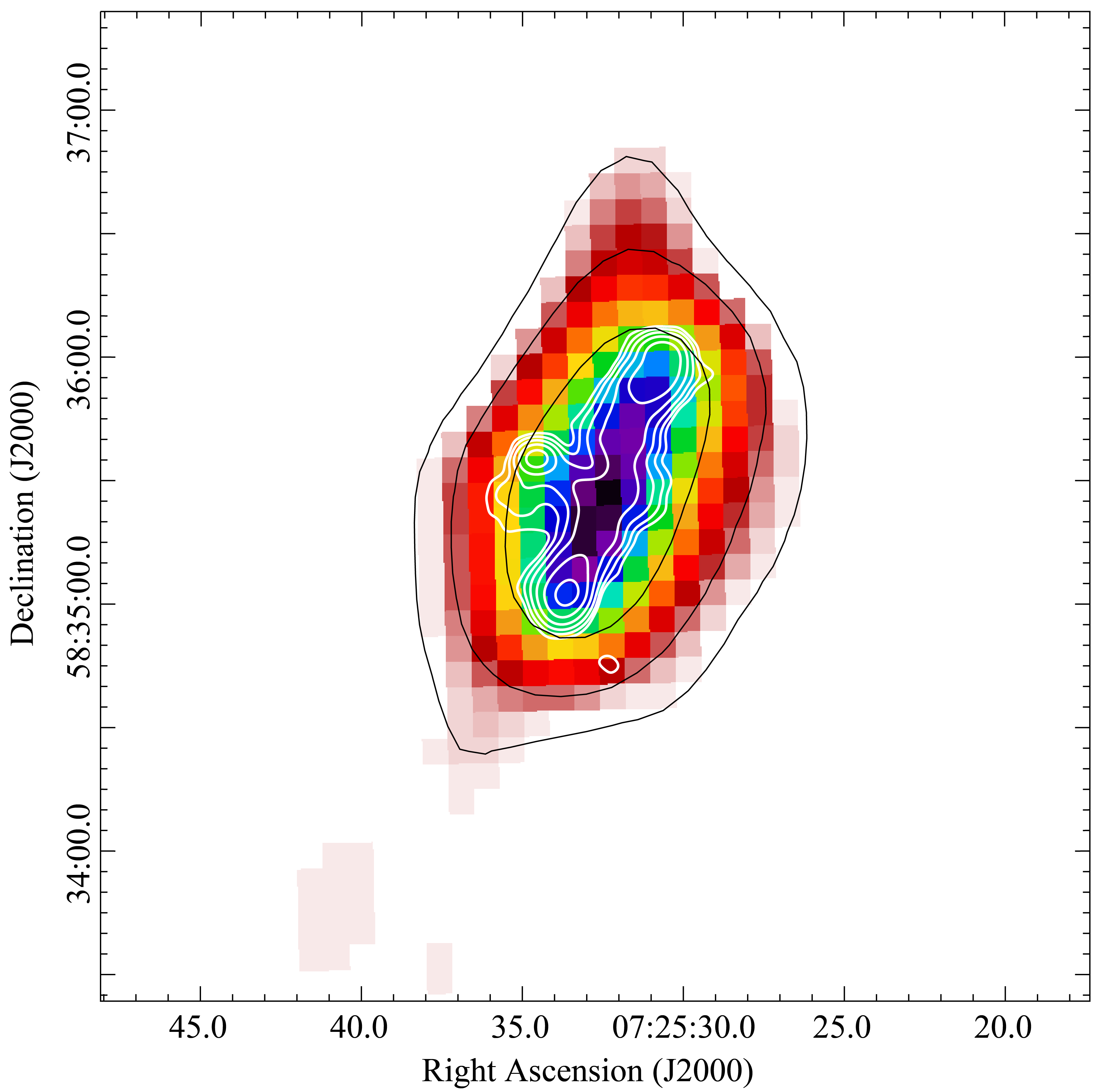}
\caption{VLA B-array L-band contours (white lines) overlaid on the TGSS 150\,MHz image and contours (black lines). Here the TGSS contours are in steps of $(0.008, 0.032, 0.128)\,\mathrm{Jy\,beam^{-1}}$, and the VLA contours are in steps of $(1,2,4,...)\times0.309\,\mathrm{mJy\,beam^{-1}}$. The pseudo-color image illustrate TGSS radio flux density distribution. \label{fig:tgss}}
\end{figure}

In the VLA S-band image (Figure \ref{fig:vla}), there are signatures supporting jet re-orientation in our target \citep[see][for a summary of signatures that supports jet re-orientation]{2019MNRAS.482..240K}, here we marked it follow \citet{2019MNRAS.482..240K}, see Figure \ref{fig:vla}: (R) ring-like structure and wide hotspot on the northern side, and the southern hotspot is not at the edge of its lobe, (S) the northern hotspot on the opposite side of lobe/jet as the southern hotspot. The signatures of jet re-orientation were seen in the jetted kiloparsec-scale \citep[NGC\,326, ][]{2001A&A...380..102M} and parsec-scale \citep[0402+379, ][]{2006ApJ...646...49R} dual/binary AGN systems. More interestingly, the VLA structure of \obj\ is similar to 0402+379 \citep{2006ApJ...646...49R} and appears as a scaled-up version. Again, lateral flow signatures were found in the spectral index map of the southern hotspot (i.e. the spectral index gradient in Figure \ref{fig:spmap}) which supports the ongoing jet re-orientation. Further evidence that supports jet precession is from 150\,MHz image (Figure \ref{fig:tgss}), where the northern jet extension is different from the VLA images.

There is no evidence, e.g. the FRI type main jets, double boomerang morphology, and bright apex in each boomerang \citep[see][]{2020MNRAS.495.1271C}, support the backflow is the main mechanism that is responsible for the X-shaped radio morphology. Furthermore, the major jet direction of core B is nearly aligned with the minor axis of the host (see Figure \ref{fig:rband} in Appendix), which also disfavors the backflow model \citep{2016AA...587A..25G}. According to the above evidence of jet distortion, we favor an explanation in terms of a fast realignment of the jet axis for the X-shaped jet in \obj. In addition, the backflowing plasma from the major jets maybe at work in assisting to form the `wings'. For example, \citet{2012A&A...545L...3C} propose that after a re-orientation the new jet backflow is deflected into the fossil cavities created in the previous active phase.

Though the association between kpc scale binary (or dumbbell) galaxies and distorted jet structure was well addressed \citep{1980A&A....85..101B, 1982AJ.....87..602W, 1991AJ....102.1960P, 1993ApJ...416..157B, 2001A&A...380..102M}, it's unlikely in XRGs that the evolution of kilo-parsec-scale galaxy pairs can drive dramatic jet realignment from wings to major jet \citep[see][]{2019MNRAS.482..240K}. Alternative models include a sudden spin-flip in galaxy merger \citep{2002Sci...297.1310M} and interaction of pc-scale binary SMBHs \citep{2004MNRAS.347.1357L}. Therefore, it is unlikely that galaxies A and B, as the pair in \obj, are responsible for the X-shaped jet. In the spin-flip model, the jet axis will only change once, it can explain the quick jet realignment from $R_1$ (the wings) to $R_2$ (major jet direction). While the deficit of the spin-flip model is that the later jet re-orientation cannot be explained. Again, we will see below \ref{subsec:merger} that an on-going merger signature appears in galaxy B, which still disfavors the spin-flip model. On the other hand, the jet precession model can explain the ongoing jet re-orientation. However the jet precession model may require pc-scale binary SMBHs, which is not seen in our VLBI observation, a possible explanation here could be its binary companion is not radio-frequency active.

\subsection{Host galaxies} \label{subsec:merger}

Taking an averaged redshift $z\sim0.42$, we can estimate black hole mass by using (a) the correlation between (5\,GHz) radio power $P_5$ and black hole mass $M_\mathrm{BH}$ \citep{2001ApJ...551L..17L}. While the $L_5 - M_\mathrm{BH}$ correlation couples with Eddington ratio \citep{2002ApJ...564..120H}, here we roughly assume an Eddington ratio of $0.1$ in our case \citep[e.g.][]{2001ApJ...551L..17L}. The 5\,GHz radio flux density for both core A and core B is measured by taking VLA A-array L-band and B-array S-band flux density. The 5\,GHz radio powers for cores A and B are $P_\mathrm{5,A}=10^{24.0}\,\mathrm{W\,Hz^{-1}}$ and $P_\mathrm{5,B}=10^{23.7}\,\mathrm{W\,Hz^{-1}}$, respectively, corresponding to a black hole mass $M_\mathrm{BH}\sim10^9\,\mathrm{M_\odot}$ \citep{2001ApJ...551L..17L} for both galaxies A and B with an intrinsic dispersion of 1 dex due to Eddington ratio uncertainties; (b) the correlation between $M_\mathrm{BH}$ and $B$-band absolute magnitude of bulge \citep[despite the host classification, we assume $B$-band magnitude of our target is bulge-dominated, ][]{2013ApJ...764..151G}. Here the Pan-STARRS1 $r$-band corresponding to a rest-frame $B$-band in our case (the effective wavelength $\lambda_{eff, B}=4400\,\mathrm{\AA}$). Therefore, the $B$-band absolute magnitude \citep[with Galactic absorption correction based on][]{2011ApJ...737..103S} for galaxies A and B are $M_{r,\mathrm{A}}=-21.41\pm0.03$ and $M_{r,\mathrm{B}}=-21.25\pm0.05$, respectively, which yield a black hole mass $M_\mathrm{BH} = 10^9 - 10^{10}\,M_\odot$ for both galaxies A and B; (c) the correlation between $M_\mathrm{BH}$ and $K$-band absolute magnitude of bulge \citep[assuming $K$-band magnitude of our target is bulge-dominated, ][]{2013ARA&A..51..511K}. Here the WISE $W1$-band corresponding to a rest-frame 2MASS $K_s$-band in our case (the effective wavelength $\lambda_{eff, K_s}=2.159\,\mu{}m$). Therefore, the $K$-band absolute magnitude \citep[with Galactic absorption corrections based on][]{2011ApJ...737..103S} for galaxies A and B are $M_{K,\mathrm{A}}=-26.54\pm0.03$ and $M_{K,\mathrm{B}}=-26.31\pm0.04$, respectively, which again yield a black hole mass $M_\mathrm{BH} = 10^9 - 10^{10}\,M_\odot$ for both galaxies A and B, here the $M_K - M_\mathrm{BH}$ correlation is tighter than (a) and (b) \citep[see][]{2013ARA&A..51..511K}. The three estimates of black hole mass are consistent with each other, which induce a total galaxy stellar mass $M_{stellar}=10^{11} - 10^{12}\,M_\odot$ \citep{2015ApJ...813...82R, 2020ARA&A..58..257G}. The stellar mass is consistent with 100\,kpc scale galaxy pairs \citep{2020ApJ...904..107S}.

The mid-infrared magnitudes for the two cores have been collected from AllWISE Data Release \citep{2014yCat.2328....0C} as $m_{W1,\mathrm{A}}=15.296\pm0.038$, $m_{W2,\mathrm{A}}=15.018\pm0.075$, $m_{W3,\mathrm{A}}=12.200$ for core A, and $m_{W1,\mathrm{B}}=15.520\pm0.041$, $m_{W2,\mathrm{B}}=15.444\pm0.097$, $m_{W3,\mathrm{B}}=12.089$ for core B, which yield the mid-infrared colors $m_{W1,\mathrm{A}}-m_{W2,\mathrm{A}}=0.27\pm0.08$ and $m_{W2,\mathrm{A}}-m_{W3,\mathrm{A}}=2.81\pm0.07$ for galaxy A, and $m_{W1,\mathrm{B}}-m_{W2,\mathrm{B}}=0.07\pm0.10$ and $m_{W2,\mathrm{B}}-m_{W3,\mathrm{B}}=3.35\pm0.09$ for galaxy B. Galaxy A is located in the transition region from early-type spirals \citep[or Sa/SBa-type in Hubble's classification scheme, with semiquiescent star formation, see descriptions in][]{2019ApJS..245...25J} to active star-forming galaxies, while galaxy B is located in the region of active star-forming galaxies \citep{2019ApJS..245...25J}. It is shown that 100\,kpc scale galaxy pairs already have tidal interaction \citep{2011ApJ...737..101L}, while it has no help on the AGN activity \citep{2020ApJ...904..107S}. Therefore, the star-forming activity in galaxy B is more likely self-induced, e.g. an ongoing merger \citep{2006ApJS..163....1H, 2007A&A...468...61D}.

\section{Summary}
We perform a high-resolution EVN 5\,GHz observation of the XRG \obj, two compact and non-thermal radio cores are detected. We have also re-processed archival VLA 1.5 and 3\,GHz data, and the radio spectral index map and astrometric coordinate have been measured in this work. Comparing with the VLA and EVN observations: (1) We find XRG \obj\ is associated with a $100$\,kpc-scale dual jetted-AGN system, where one of them is Faranoff-Riley type I (FR I), the other one is FR II; (2) We find evidence that the jet in core B is changing its direction, which naturally forms an X-shaped structure in \obj. Two models are discussed on the origin of jet re-orientation, including the sudden flip of black hole spin and jet precession due to the interaction of parsec-scale binary SMBHs. Future low-frequency radio observations with high sensitivity may help to reveal the fossil radio emission; furthermore, obtaining a spectroscopic redshift and high signal-to-noise ratio image in optical is crucial to identify the proposed mechanisms in this work.

\begin{acknowledgments}
This work is supported by the National Science Foundation of China (12103076) and the National Key R\&D Programme of China (2016YFA0400702, 2018YFA0404602, 2018YFA0404603). 
XLY is supported by the Shanghai Sailing Program (21YF1455300) and China Postdoctoral Science Foundation (2021M693267).
LCH was supported by the National Science Foundation of China (11721303, 11991052). 
XLY and TA thank the financial support of the Bureau of International Cooperation, Chinese Academy of Sciences (114231KYSB20170003). 
% Scientific data
Scientific results from data presented in this publication are derived from the EVN project RSY07 and VLA project 16A-220 and 16B-023.
% EVN
The European VLBI Network (EVN) is a joint facility of independent European, African, Asian, and North American radio astronomy institutes.
% China SKA Regional Centre prototype
The VLBI data processing in this work made use of the compute resource of the China SKA Regional Centre prototype, funded by the Ministry of Science and Technology of China and the Chinese Academy of Sciences. 
% VLA
This work made use of Karl G.\ Jansky Very Large Array (VLA) data. The National Radio Astronomy Observatory is a facility of the U.S. National Science Foundation operated under cooperative agreement by Associated Universities, Inc.
% TGSS
This work made use of TGSS data. We thank the staff of the GMRT that made these observations possible. GMRT is run by the National Centre for Radio Astrophysics of the Tata Institute of Fundamental Research.
% DSS POSS-II (removed)
%\end{acknowledgments}
%\begin{acknowledgments}
% AllWISE
This research has made use of the NASA/IPAC Infrared Science Archive, which is funded by the National Aeronautics and Space Administration and operated by the California Institute of Technology. 
% DECaLS
The Photometric Redshifts for the Legacy Surveys (PRLS) catalog used in this paper was produced thanks to funding from the U.S. Department of Energy Office of Science, Office of High Energy Physics via grant DE-SC0007914.
% Pan-STARRS
This work made use of data from the Pan-STARRS1 Surveys (PS1).
\end{acknowledgments}

\appendix

\section{Measuring the orientation of galaxy B}
\begin{figure}
\centering \includegraphics[scale=0.5]{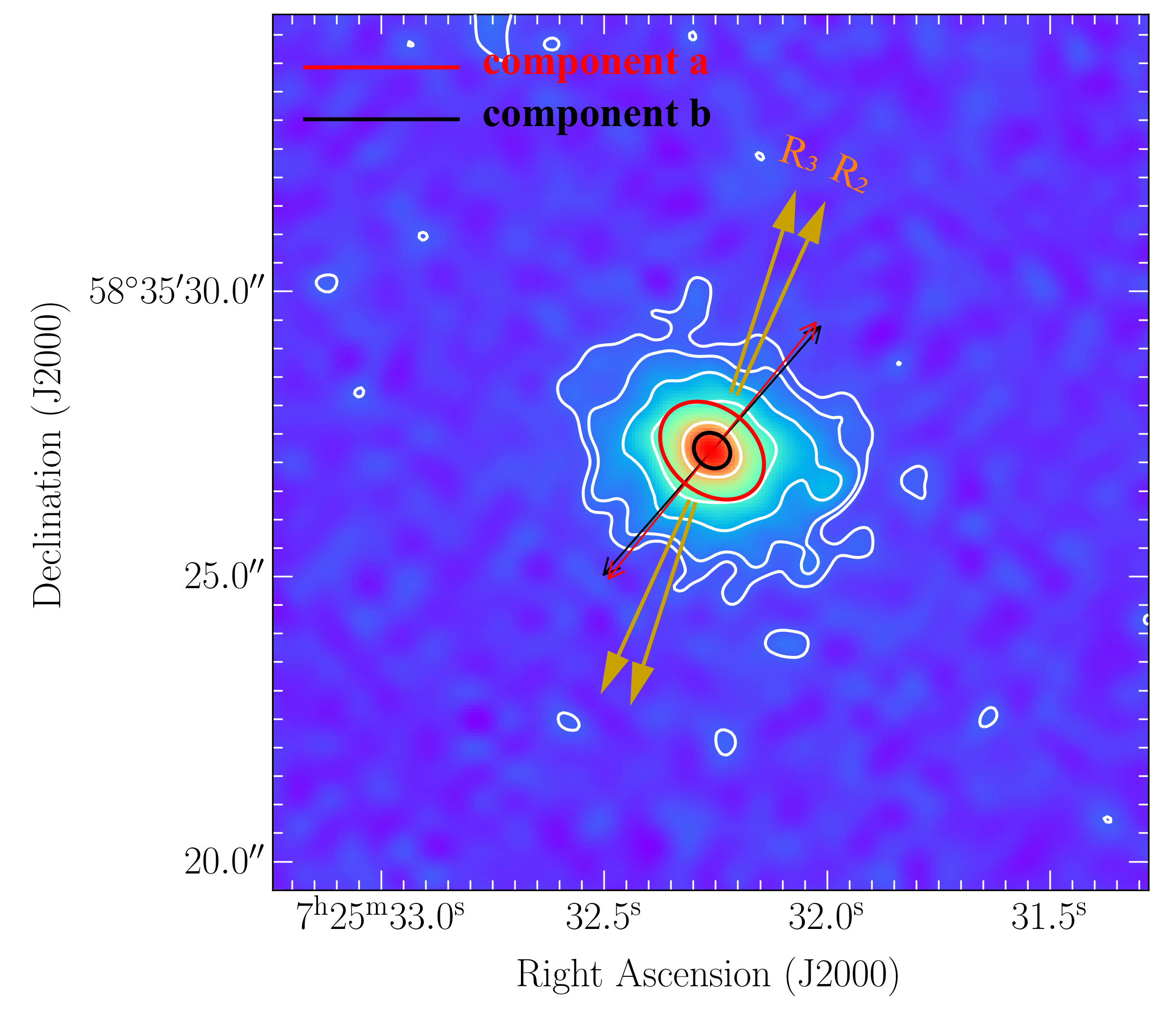}
\caption{The DECaLS \citep{2019AJ....157..168D} $r$-band image and contours (white lines) of galaxy B. $R_2$ and $R_3$ are obtained from Figure \ref{fig:model}. The black and red ellipses indicate the FWHMs of two dimensional Gaussian components that are fitted to the $r$-band image, where the black and red arrows are their minor axis directions, respectively. \label{fig:rband}}
\end{figure}

\begin{figure}
\centering \includegraphics[scale=0.35]{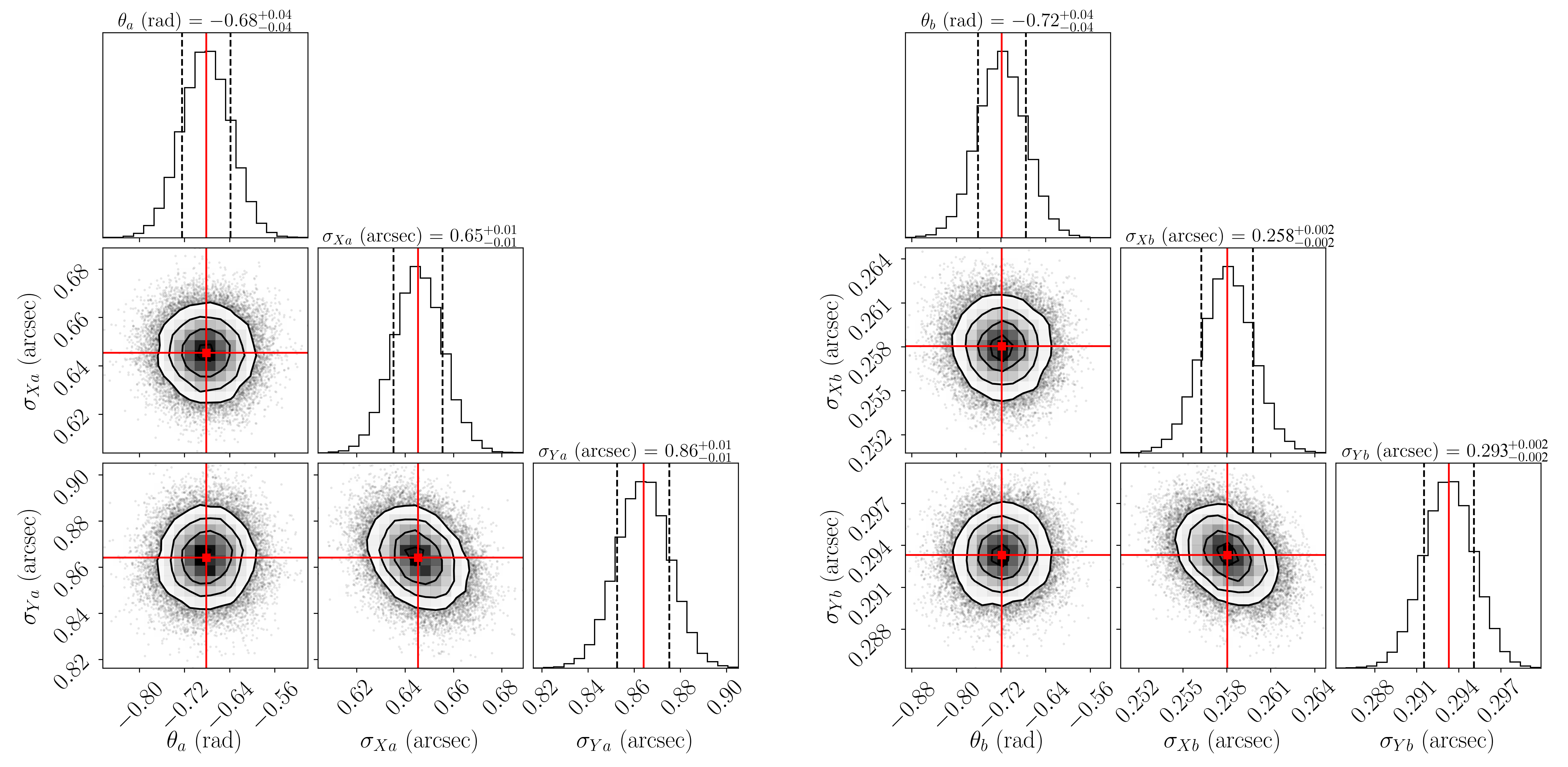}
\caption{Marginalized and joint posterior probability distribution of the key parameter values obtained from the MCMC fitting. The left and right panels show parameters for the components `a' and `b', respectively (see also Figure \ref{fig:rband}). The histograms on the diagonal show the marginalized posterior densities for each parameter, i.e. position angles ($\theta$, in radians) and standard deviation (in arcsec) of the major ($\sigma_Y$) and minor ($\sigma_X$). We take 50\% of the distributions as the best-fit values, which are marked in red cross-hairs and red lines. The uncertainties are computed as the 16\% and 84\% of the distributions, thus representing $1\sigma$ confidence ranges shown with black dashed lines. \label{fig:mcmc}}
\end{figure}

To examine the major and minor axis directions of the optical host of galaxy B, we obtained the $r$-band image from the DECam Legacy Survey \citep[DECaLS,][]{2019AJ....157..168D}. Two-dimensional elliptical Gaussian models are fitted to the DECaLS $r$-band image through the Markov chain Monte Carlo algorithm. The image can be fitted with two elliptical Gaussian components, which are designated as components `a' and `b'. The fitted elliptical Gaussian models are plotted in Figure \ref{fig:rband}, where the ellipses indicate the full width at half maximum (FWHM) along the major and minor axes. In Figure \ref{fig:mcmc}, we show the posterior probability distributions of the position angles of the major axes and the standard deviations (${\sim}\frac{FWHM}{2.355}$) of the major and minor axes. We take 16\% and 84\% of the distributions as the lower and upper limits, respectively.

%\section{Rotating tables} \label{sec:rotate}

%The process of rotating tables into landscape mode is slightly different in
%\aastex v6.31. Instead of the {\tt\string\rotate} command, a new environment
%has been created to handle this task. To place a single page table in a
%landscape mode start the table portion with
%{\tt\string\begin\{rotatetable\}} and end with
%{\tt\string\end\{rotatetable\}}.

%Tables that exceed a print page take a slightly different environment since
%both rotation and long table printing are required. In these cases start
%with {\tt\string\begin\{longrotatetable\}} and end with
%{\tt\string\end\{longrotatetable\}}. Table \ref{chartable} is an
%example of a multi-page, rotated table. The {\tt\string\movetabledown}
%command can be used to help center extremely wide, landscape tables. The
%command {\tt\string\movetabledown=1in} will move any rotated table down 1
%inch. 


\clearpage
\bibliography{J0725}{}
\bibliographystyle{aasjournal}



\end{document}